# A Broadband Dipolar Resonance in THz Metamaterials

Bagvanth Reddy Sangala, Harshad Surdi, Achanta Venu Gopal, and S. S. Prabhu

*Department of Condensed Matter Physics, Tata Institute of Fundamental Research, Mumbai, India*

Email id: *s.reddy@tifr.res.in*

**Abstract:** We demonstrate a THz metamaterial with broadband dipole resonance originating due to the hybridization of LC resonances. The structure optimized by finite element method simulations is fabricated by electron beam lithography and characterized by terahertz time-domain spectroscopy. Numerically, we found that when two LC metamaterial resonators are brought together, an electric dipole resonance arises in addition to the LC resonances. We observed a strong dependence of the width of these resonances on the separation between the resonators. This dependence can be explained based on series and parallel RLC circuit analogies. The broadband dipole resonance appears when both the resonators are fused together. The metamaterial has a stopband with FWHM of 0.47 THz centered at 1.12 THz. The experimentally measured band features are in reasonable agreement with the simulated ones. The experimental power extinction ratio of THz in the stopbands is found to be 15 dB.

## 1. Introduction

Metamaterials are designed composites of periodic metallo-dielectric structures that show specific properties to radiation or sound due to the dimensions and arrangement of the inclusions rather than the physical properties of the constituent metal or the dielectric [**1, 2**]. These materials were shown to exhibit interesting properties like negative refraction [**3, 4**], hyper resolution [**5, 6**], cloaking [**7**], perfect absorption [**8**], optical nonlinearity [**9, 10**], chirality [**11**], etc. Metamaterials were realized in almost all wavelength ranges, like, UV [**12**], visible [13], IR [**14**], terahertz (THz) [**15**], microwaves [**3, 7**], and radio frequencies [**16**]. Several optical components like lenses [**17**], filters [**18**], polarizers [**19**], waveplates [**20**], phase modulator [**21**], etc were demonstrated successfully in metamaterials.

Research in THz part of the electromagnetic spectrum was limited due to lack of efficient sources and detectors until the invention of terahertz time-domain spectroscopy (THz-TDS) [**22**]. Metamaterials are the best choice for designing optical components in the THz spectral region since the natural materials have poor electric or magnetic response to THz radiation. For spectroscopic applications, one would like to have filters to select required frequency range. For example, a stopband filter is useful to cut down a part of the frequency spectrum to selectively excite or de-excite a resonance mode of a sample. There have been many THz broadband stopband filters driven mainly by three kinds of approaches. The first one is, concentric nesting of the LC resonators, where LC resonators are connected physically resulting in a broadband filter [**23, 24, 25**]. The second one is by taking unit cells with nearly resonant LC resonances, in which the resulting linewidth looks broad because of excitation of many individual resonances. The third one is stacking of metamaterial layers, in which different layers having nearby LC resonance frequency will get excited resulting in a broad stopband in the transmission [**26, 27, 28, 29, 30, 31, 32**]. In this paper, we demonstrate a broadband THz metamaterial filter by hybridization of two metamaterial elements with LC resonances [**33**]. The hybridization of two LC resonant metamaterials with a gap between them results in an LC resonance and an electric dipole resonance. We find that the dipole resonance gets broader as the distance between the metamaterial elements is decreased while the LC resonance gets narrower. We fabricated the optimum metamaterial structure which showed a broad stopband. The structure has periodic hollow-T shapes made in gold pads on semi-insulating Gallium Arsenide (SI-GaAs) substrate. We simulated this metamaterial using the Radio Frequency module of COMSOL multiphysics, fabricated using electron beam lithography and studied its transmission by THz-TDS. The full width at half maxima (FWHM) of this structure is better than other single layered metamaterials reported. This paper is organized as follows, in section 2, we describe the approach for line broadening which involves by hybridization of two metamaterials (top-C and bottom-C) with LC resonances. Section 3 summarizes the optimized broadband metamaterial, its fabrication, and characterization by THz-TDS.

## 2. Hybridization of two metamaterial elements with LC resonances

Although Pendry[1] used theoretically three dimensional arrays of split ring resonators or wire arrays to get metamaterial response, for practical reasons most of the experimental realizations used two dimensional array of the unit cells consisting of split ring resonators or other structures. We used Radio Frequency module of COMSOL multiphysics to simulate THz propagation through two dimensional metamaterials. We considered metamaterials with unit cells having one (top-C or bottom-C) and two elements (two elements hybridized by a gap g). The top-C and bottom-C resonators, which have one LC resonance each, when hybridized, show an LC and an additional electric dipole resonance. As the gap between the top-C and bottom-C is decreased, the LC mode width decreases and the electric dipole width increases. Figure 1(a)-1(c) show the schematic of the unit cells. Here, c = 15μm, b = 36μm, lw = 4μm, and P = 50μm. In the simulations, we launch THz radiation normal to the plane of the metamaterials by port boundary conditions and apply periodic boundary conditions along transverse directions. Figure 1(d)-1(e) show the electric field norm and current density of the top-C and bottom-C metamaterials at their resonances. Figure 1(f) and (1g) show the LC and electrical dipole nature of the resonances of the hybrid metamaterial with g=0.5 um. Figure 1(i)-1(j) show the normalized $|S21|^2$ versus frequency of top-C and bottom-C metamaterials. Figure 1(k) and 1(l) show the normalized $|S21|^2$ versus frequency of LC and ED modes of the hybrid metamaterial for various gaps. Figure 2(b) shows the center frequency versus gap of the hybrid metamaterial. We can see the center frequency of both the modes increasing as the gap (g) is increased. This is similar to the trend of an LCR resonator. Also, in a series LCR resonator, the damping factor, which is proportional to the FWHM of the resonance is proportional to square root of capacitance and if the capacitance is modeled to be that of a parallel plate one, it is proportional to inverse of square root of the gap between the plates. In the parallel LCR resonator, the damping factor is inversely proportional to the capacitance and directly proportional to the square root of the gap between the plates. Figure 2(a) shows the full width at half maxima of the LC and ED modes for various gaps of the hybrid metamaterial. From these plots, we can say that the LC mode trend is similar to that of a parallel LCR resonator and ED mode trend is similar to that of a series LCR resonator.

## 3. A broadband THz dipole metamaterial and experimental realization

The unit cell of the hybridized metamaterial with no gap between the elements which showed broadband dipole resonance is shown in Figure 3(a). Figure 3(b) shows an SEM image of the fabricated metamaterial. Figure 3(c) shows a plot of THz waveforms transmitted through a Nitrogen ambience (reference in blue) and through the metamaterial (red line).

The metamaterial was fabricated on a semi-insulating GaAs substrate using electron-beam lithography. A positive electron beam resist (PMMA) was spin coated on a pre-baked GaAs wafer to obtain a uniform layer of approximately 200 nm thickness. An array of the unit cells was written spanning an area of 5 x 5 $mm^2$ by e-beam. The e-beam written sample is developed in MIBK IPA followed by rinsing with UV grade IPA to get a patterned layer of resist on SI-GaAs wafer. DC sputtering was done to deposit a 150 nm layer of gold on the patterned sample. The final metamaterial structure was obtained using a lift-off in acetone bath with sonication process. Residual acetone was removed from the sample surface by oxygen plasma.

The fabricated metamaterial was characterized by a THz-TDS setup in transmission mode. A photo-conductive antenna is used as the THz source. A pair of off-axis parabolic mirrors were used to focus THz to a spot of 2 mm diameter on the sample. The transmitted THz waveforms were detected by electro-optic sampling method using ZnTe crystal, quarter wave plate (QWP), Wollaston prism, and balanced photodiodes. We acquired waveforms transmitted before mounting the sample (reference scan) and after mounting the metamaterial (sample scan). All the waveforms were taken in a Nitrogen purged environment. We truncated both the waveforms at a delay time before the echo from the GaAs substrate to avoid the interference in the spectral domain. We interpolated the waveform data using *spline* function of Origin before calculating their spectra. Since 26 ps long waveforms were used for calculation of spectra, the spectral resolution was 38 GHz. We calculated transmission coefficient as the ratio of sample spectrum to the reference spectrum. Figure 4 shows the square of amplitude of transmission coefficient of the metamaterial versus frequency. The figure also shows the normalized $|S21|^2$ data from the simulations. We see a very good agreement between the transmission data of experiment and simulation. The experimental FWHM of the stopband centered at 1.12 THz is 470 GHz. Table 1 shows the comparison of the THz stopband filters reported so far. The FWHM of our metamaterials is better than other single layer counterparts and at the same time the quality factor is comparable to the highest values reported in multilayer structures.

| Unit cell type | $\nu_o$ [THz] | FWHM [GHz] | Q | Reference |
|---|---|---|---|---|
| **Nested square-C** | 0.74 | 443 | 1.67 | [24] |
| **Nested square-C Four loops** | 0.68 | 280 | 2.43 | [23] |
| **5-layered square-C** | 0.5 | 380 | 1.32 | [27] |
| **Modified square SRR** | 0.515 | 201 | 2.56 | [25] |
| **Square hole patterned Metal-Dielectric-square hole patterned Metal** | 1.26 | 350 | 3.6 | [28] |
| **Bilayer-fish-scale** | 0.7 | 1130(3dB) | 0.62 | [29] |
| **Polyamide-metal-polyamide-metal-polyamide square+cross** | 0.8 | 690(3dB) | 1.16 | [26] |
| **Double layer metal holes** | 0.8 | 400 | 2.00 | [30] |
| **Swiss cross hole patterned Metal-Dielectric-Swiss cross hole patterned Metal** | 1.25 | 500 | 2.50 | [31] |
| **Multilayer loop resonators (square+circular)** | 1.75 | 1400 | 1.25 | [32] |
| **Hollow-T gold pads on GaAs** | 1.12 | 470 | 2.38 | This work |

Table 1. Comparison of broadband THz filters reported so far.
.

**4. Summary**

We have demonstrated, by numerical simulations, line broadening due to hybridization of two LC resonators in a metamaterial. We realized experimentally, a broad stopband filter consisting of an array of hollow-T gold pads on SI-GaAs. The metamaterial has a stopband with FWHM of 470 GHz and center frequency of 1.12 THz. The power extinction ratio of THz at the center of the stopband is experimentally found to be 15 dB.

**Figure Captions**

**Figure1**. (a)-(c). Geometry of top-C, bottom-C, and their hybrid. Here, P=50 μm, b=36 μm, c=15 μm, lw=4μm, g∈[0, 4]um, and Au pads height on SI-GaAs wafer is 150 nm. (d)-(g) Show electric field norm (color bar) and current density direction(red arrows) of the metamaterials. (f) Shows the LC nature of the first hybrid resonant mode and (g) shows the electrical dipole (ED) nature of the second mode. (e)-(h) Show normalized $|S21|^2$ versus frequency of top-C and bottom-C metamaterials. (i)-(j). Normalized $|S21|^2$ data of top-C and bottom-C metamaterials. (k)-(l) Show normalized $|S21|^2$ data of LC and ED modes for various gaps of hybrid. The red arrow lines between (a) and (b) show polarization of incident THz radiation.

**Figure 2**. (a) Shows the full width at half maxima versus gap and (b) shows center frequency versus gap of the hybrid metamaterial for LC and ED modes.

**Figure 3**. (a) Unit cell of the metamaterial. Here, b=36 μm, c=15 μm, lw=4 μm, and P=50 μm, and height of the gold pads on SI-GaAs is 150 nm. (b) An SEM image of the fabricated metamaterial. (c) The THz waveforms transmitted through a reference and the metamaterial sample.

**Figure 4**. Shows the simulated normalized $|S21|^2$ and experimental power transmission coefficient versus frequency.

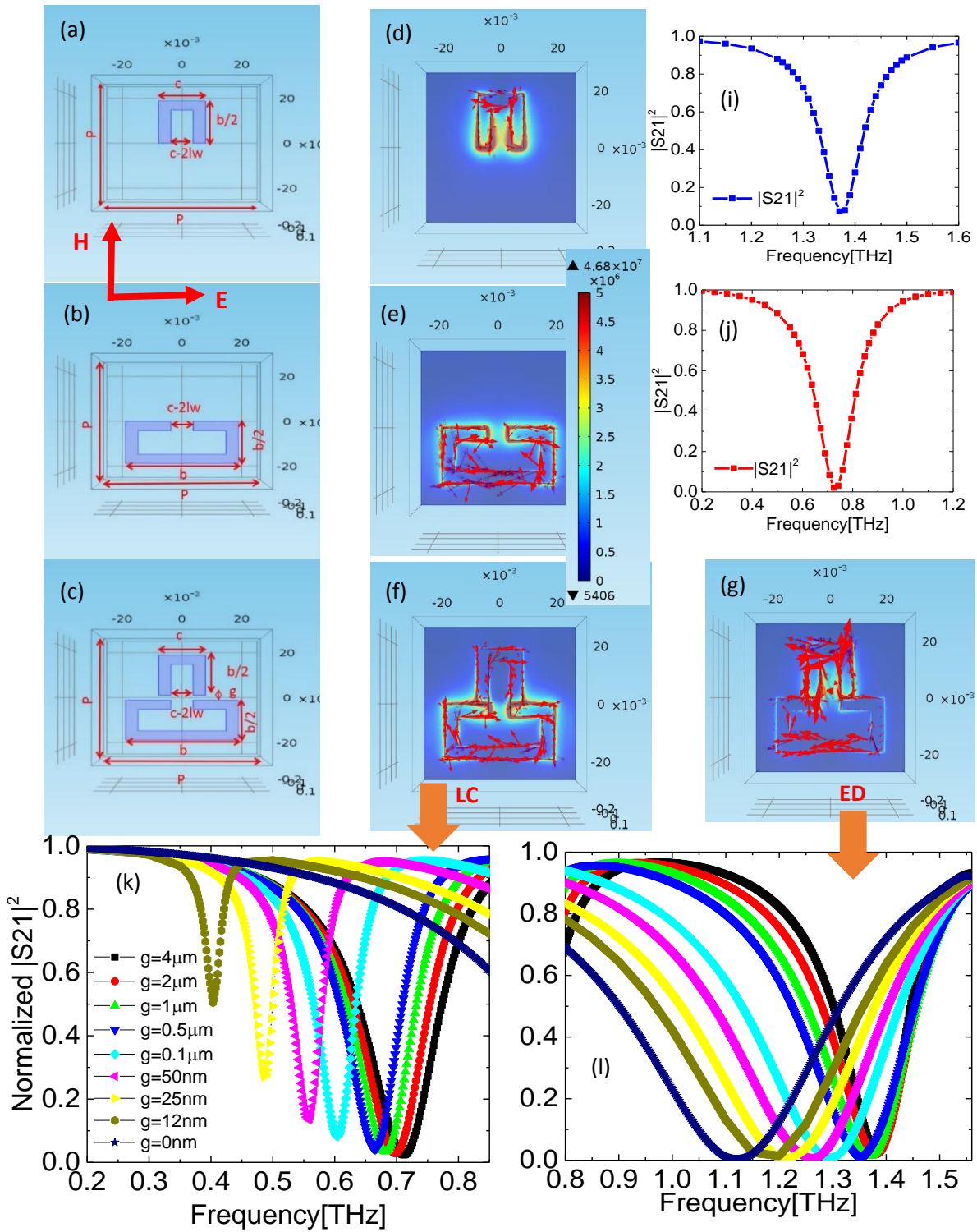

**Figure1.**

Bagvanth R. Sangala *et al*.

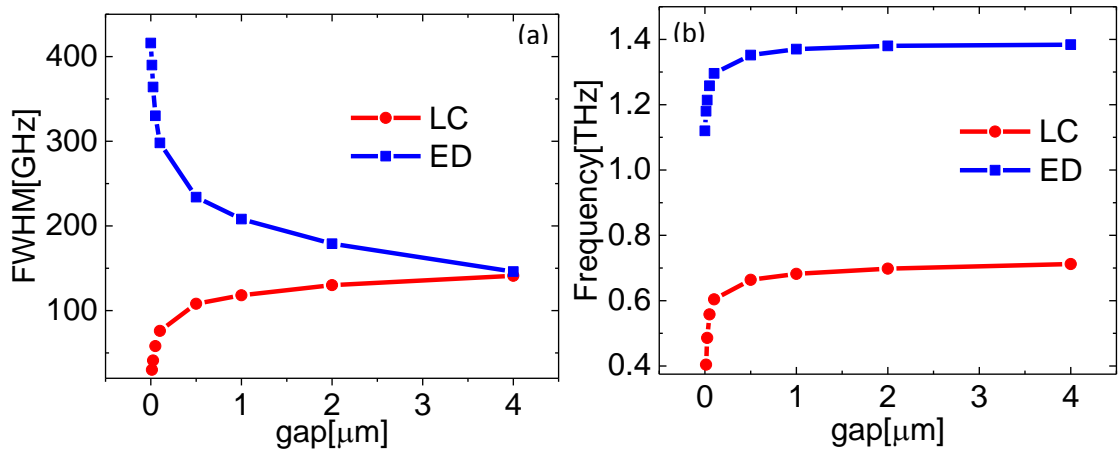

**Figure2.**

**Bagvanth R. Sangala** *et al*.

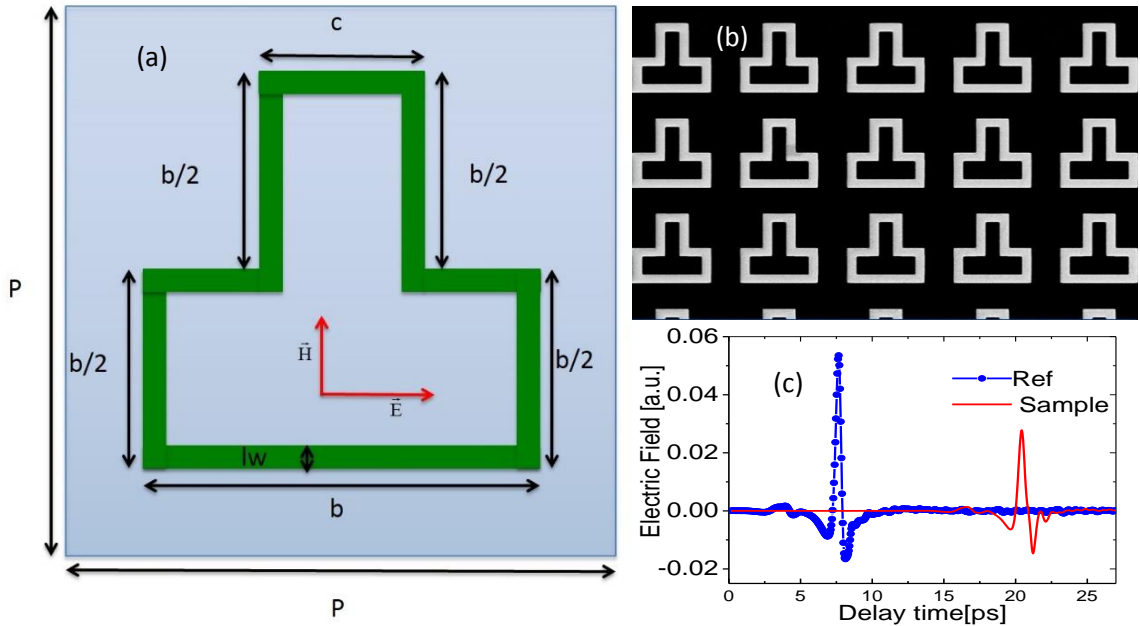

**Figure3.**

**Bagvanth R. Sangala** *et al*.

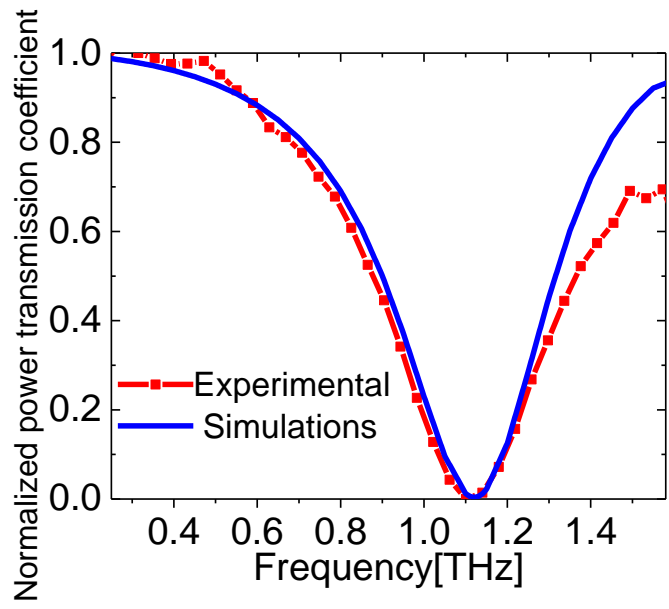

**Figure4.**

**Bagvanth R. Sangala** *et al*.